# Predicting Stable Phase Monolayer Mo$_2$C (MXene), a Superconductor with Chemically-Tunable Critical Temperature

*Jincheng Lei, Alex Kutana, and Boris I. Yakobson\**

Department of Materials Science and NanoEngineering, Rice University, Houston, Texas 77005, United States

*E-mail: biy@rice.edu.

## Abstract

Two-dimensional (2D) superconductors have attracted great attention in recent years due to the possibility of new phenomena in lower dimensions. With many bulk transition metal carbides being well-known conventional superconductors, here we perform first-principles calculations to evaluate the possible superconductivity in a 2D monolayer Mo$_2$C. Three candidate structures (monolayer alpha-Mo$_2$C, 1T MXene-Mo$_2$C, and 2H MXene-Mo$_2$C) are considered and the most stable form is found to be the 2H MXene-Mo$_2$C. Electronic structure calculations indicate that both unpassivated and passivated 2H forms exhibit metallic properties. We obtain the phonon frequencies and electron-phonon couplings using density-functional perturbation theory, and based on the BCS theory and McMillan equation, estimate the critical temperatures to be in the ~0-13 K range, depending on the species of the surface termination (O, H and OH). The most interesting termination group is H, which can increase the electron-phonon coupling and bring the critical temperature to 13 K. This shows a rather high critical temperature, tunable by surface termination, making this 2D carbide an interesting test bed for low-dimensional superconductivity.

## Introduction

Superconductivity in 2D materials has been a focus of recent intensive research. Previously, superconductivity was explored theoretically in doped graphane,[1] graphene,[2] B$_2$C single layers,[3] doped phosphorene,[4] and, most recently, two-dimensional boron.[5,6] However, graphane, graphene and phosphorene require substantial doping to achieve good metallicity, whereas other intrinsically metallic materials such as B$_2$C or 2D boron are either hypothetical or may have their properties strongly affected by the growth substrate. In contrast, MXenes, or monolayer transition metal carbides and carbonitrides, have been synthesized as standalone 2D layers with measured good metallic conductivity.[7] Additionally, bulk carbides, especially transition metal carbides, have been known for years as superconductors.[8]

Two-dimensional transition metal carbides (TMCs) have gained great attention in recent years as their 2D crystals have been grown by chemical vapor deposition (CVD), and also obtained through hydrofluoric acid etching of MAX phases. Recently, Xu *et al.*[9,10] reported the chemical vapor deposition (CVD) synthesis of large-scale, high-quality, ultrathin alpha-$Mo_2C$ 2D crystals and demonstrated suppression of the superconducting temperature with decreasing thickness.[9] Unfortunately, great challenges still exist in CVD growth of $Mo_2C$ monolayer samples. At the same time, monolayer 2D transition metal carbides known as MXenes can be produced by selectively etching "A" layers out from the parent MAX bulk phase.[7] Here, M designates an early transition metal, A represents group A elements, and X is C and/or N. Thus synthesized MXenes are always terminated with functional groups, like hydroxyl and oxygen.[11] Since 2011, a variety of MXenes have been successfully synthesized with this method.[12-15] The exceptional combination of TMC core and surface terminating groups endows MXenes with good metallic conductivity and surface hydrophilicity, as well as outstanding properties[16] for applications in electronic devices,[17] gas sensors,[18] water purification,[19] energy storage,[20] *etc.*

Unlike few-layer alpha-$Mo_2C$ crystals, superconductivity has not yet been demonstrated in monolayer MXenes, and it is still unclear whether the superconducting state could survive as the thickness is reduced to a monolayer. MXenes may thus be a good test bed for exploring superconductivity in monolayer transition metal carbides and some of the best candidates to display superconductivity among other 2D materials, provided that other important prerequisites for the BCS superconductivity are met. This prompts us to theoretically explore conventional superconductivity of MXenes, as described by the BCS theory.[21]

Quite recently, $Mo_2C$ MXene has been exfoliated from bulk $Mo_2Ga_2C$,[22,23] a unique MAX phase with two "A" layers.[24] In this paper, we identified the most stable form of the monolayer $Mo_2C$ and established the relationship between bulk alpha-$Mo_2C$ and the monolayer. Given that surface terminations can change the relative stability of different phases of a 2D material, the functionalized derivatives of $Mo_2C$ monolayer were studied as well. Basic band structure calculation indicates that all the considered monolayers are metals, meeting the prerequisite for superconductivity. The superconducting critical temperatures ($T_c$) of bulk alpha-$Mo_2C$ and monolayer counterpart have been estimated, and agree with the reported reduction of $T_c$ with thickness in $Mo_2C$. The variations of $T_c$ with surface termination functional groups in $Mo_2C$ monolayers were further explored. The rather high and well-tunable $T_c$ makes $Mo_2C$ monolayer an appealing superconductor.

# Results and discussion

As a starting point, we focused on the atomic structure of monolayer $Mo_2C$. Halim *et al.*[23] carefully characterized this new member of the MXene family, but the exact atomic arrangement is still controversial: Khazaei[25] and Sun[26] have separately put forward two different atomic structures, which we designate as MXene 1T and 2H phases, in keeping with transition metal dichalcogenides (TMDs) notation.[27] In addition, the CVD growth may potentially produce monolayer alpha-$Mo_2C$. Having these in mind, we built three different $Mo_2C$ monolayer structures: monolayer alpha-$Mo_2C$, 1T MXene-$Mo_2C$, and 2H MXene-$Mo_2C$. The monolayer alpha-$Mo_2C$ was constructed by truncating the 3D bulk phase[28] perpendicular to the [001] direction, as shown in Fig. 1a. The structure obtained by truncating normal to other low-index direction is however higher in energy or cannot exist as a monolayer (see Fig. S1), and will not be discussed further.

The relaxed alpha-$Mo_2C$ monolayer has an orthorhombic unit cell, as shown in Fig. 1b. The equilibrium lattice parameters are found to be $a = 2.70$ Å and $b = 5.86$ Å. In contrast, the two types of MXene-$Mo_2C$ monolayers are arranged in hexagonal lattices. Atoms stack in a triple layer in a sequence of Mo−C−Mo. For the 1T phase, two Mo atomic layers stack in the A-B packing mode, whereas in the 2H phase (see Fig. 1c), one Mo layer shifts, turning the packing into the A-A mode. More details on the lattice constants and total energies of these structures are given in Table 1.

**Table 1** Lattice constants and relative total energies $E$ (eV per $Mo_2C$ unit) of $Mo_2C$ monolayers. The lowest energy is used as the reference.

| Monolayer $Mo_2C$ | $a$ (Å) | $b$ (Å) | $E$ (eV) |
|---|---|---|---|
| Alpha phase | 2.70 | 5.86 | 2.03 |
| 1T MXene | 2.88 | / | 0.25 |
| 2H MXene | 2.82 | / | 0 |

The structural stability of the three possible structures (monolayer alpha-$Mo_2C$, 1T MXene-$Mo_2C$, and 2H MXene-$Mo_2C$) was estimated by comparing their relative total energies. Clearly, the 2H MXene-$Mo_2C$ is the most stable one among the three. Therefore, in the monolayer form, the 2H structure is most likely to be observed. The monolayer alpha-$Mo_2C$ is very likely to convert into the MXene-$Mo_2C$ due to the considerable differences in energy. Fig. 1b shows one possible pathway for the alpha-$Mo_2C$ monolayer to transform into the lowest-energy 2H MXene-$Mo_2C$ (the initial unit cell in Fig. 1b transforms into the cell shown with the red dotted lines in Fig. 1c). However, further experimental and theoretical studies are needed to confirm this pathway. In view of the large energy excess of monolayer alpha-$Mo_2C$ relative to MXene-$Mo_2C$ structures and consequently high chance of spontaneous transformation from monolayer alpha-$Mo_2C$ to MXene-$Mo_2C$, hereafter, we will mainly discuss the MXene-$Mo_2C$ monolayers.

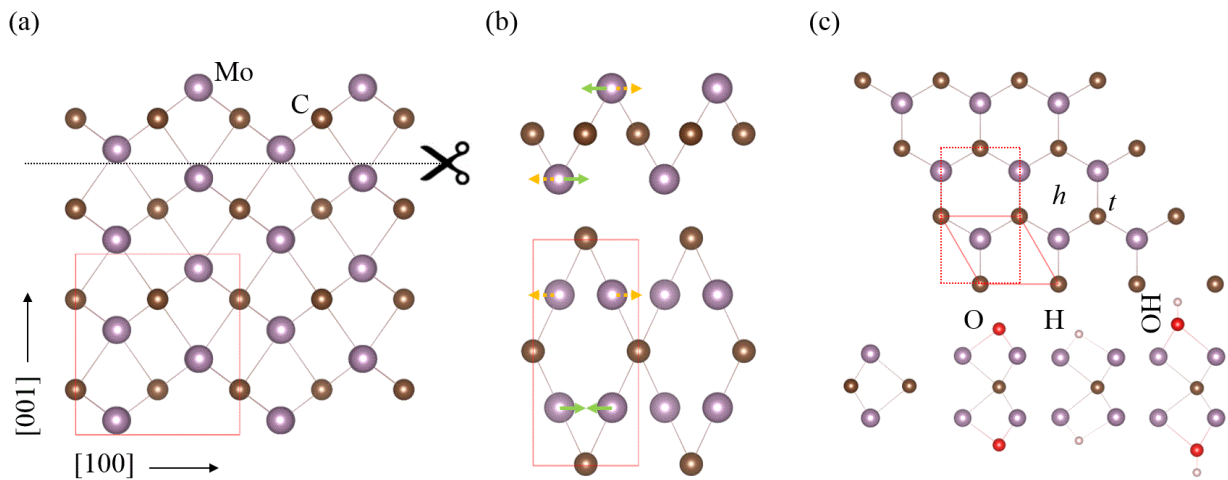

**Fig. 1** Optimized geometries of (a) alpha-$Mo_2C$, (b) monolayer $Mo_2C$ obtained by truncating from the alpha phase (side and top view), and (c) 2H MXene-$Mo_2C$ (top and side view) and its derivatives (side view). The $h$ and $t$ represent the possible adsorption sites for surface terminations. The red solid lines exhibit the unit cells.

Note that the energy difference between the 1T and 2H phases of MXene-$Mo_2C$ is 0.25 eV per $Mo_2C$ unit, which is much less than that of $MoS_2$, where the difference is 0.84 eV per $MoS_2$ unit.[27] We therefore considered the influence of surface functionalization on the stability of MXene-$Mo_2C$ monolayer, since surface terminations can change the relative stability of different phases of monolayer $MoS_2$.[29] The corresponding derivatives were constructed by saturating the surface Mo atoms, which are under-coordinated. Khazaei *et al.*[25] have investigated the possible functionalization of 1T MXene-$Mo_2C$ with O and OH groups using DFT calculations. According to their study, the most stable configuration for 1T-$Mo_2CO_2$ is one where all of the O atoms are located above the topmost sites of C atoms. In contrast, all the OH terminations in 1T-$Mo_2C(OH)_2$ prefer the hollow sites between the three adjacent C atoms. Similar to OH groups, Hui Pan found[30] that H atoms in 1T-$Mo_2CH_2$ tend to be positioned above the hollow sites. Hence, for the 1T MXene-$Mo_2C$ monolayer we only focus on these most stable configurations.

To the best of our knowledge, no previous work reported the functionalization of 2H MXene phase, including $Mo_2C$. Here, for the first time, we systematically studied the functionalized derivatives of 2H MXene-$Mo_2C$ monolayer. In accordance with normal practice,[31] we considered three different arrangements of functional groups on the surface of the 2H phase (see Fig. 1c): I) all functional groups are placed above the hollow sites (site $h$); II) all functional groups are situated on the topmost sites of C atoms (site $t$), and III) on one side, all the functional groups locate above the hollow sites, while on the other side, all the functional groups are at the topmost sites. It turns out that in 2H MXene-$Mo_2C$ monolayer, the H atoms still favor the hollow sites (model I), and O atoms are still most likely to stay at the topmost sites of C atoms (model II). However, the OH groups now tend to adopt an asymmetric arrangement, *i.e.* above the hollow sites on one side while at the topmost sites on the other side (model III).

Table 2 summarizes the computed lattice constants and relative total energies of the functionalized MXene-$Mo_2C$ monolayer derivatives of the most stable 1T structures and the three possible 2H configurations. A general tendency of the lattice constant is that it decreases from the 1T phase to the 2H phase. The adsorption of functional groups on the surface generally results in the lattice extension, except for 1T-$Mo_2CO_2$, which decreases by 1%. Comparing total energies, we found that all the 2H configurations are lower in energy than the corresponding 1T structures, suggesting that the 2H phase MXene-$Mo_2C$ is more stable than the 1T phase. This is in contrast with monolayer $MoS_2$, whose stability depends on the surface functionalization.[29] Note that we also tried to functionalize a single layer of the bulk alpha phase of $Mo_2C$, and found that the initial structures changed dramatically upon relaxation, with their final energies being still higher than those of the corresponding 2H structures. The 2H phase MXene-$Mo_2C$ is thus most energetically stable, regardless of the existence and type of the functional group. Note that $Mo_2C$ is just one representative of the large MXene family. Herein we would like to call attentions to the possibility of the 2H phase in other MXenes. Because of the difference in the atomic arrangements, substantially different properties may result, as seen below in the electronic properties of $Mo_2C$ MXene.

**Table 2** Lattice constants $a$ (Å) and relative total energies $E$ (eV per $Mo_2C$ unit) of 2H MXene-$Mo_2C$ functionalized derivatives. The lowest energies are used as the references.

| MXene derivative | 1T Phase | | 2H Phase | | | | | |
| --- | --- | --- | --- | --- | --- | --- | --- | --- |
| | | | Model I | | Model II | | Model III | |
| | $a$ (Å) | $E$ (eV) | $a$ (Å) | $E$ (eV) | $a$ (Å) | $E$ (eV) | $a$ (Å) | $E$ (eV) |
| $Mo_2CO_2$ | 2.85 | 0.39 | 2.87 | 1.03 | 2.85 | 0 | 2.87 | 0.65 |
| $Mo_2CH_2$ | 2.90 | 0.26 | 2.88 | 0 | 2.87 | 0.55 | 2.88 | 0.23 |
| $Mo_2C(OH)_2$ | 3.24 | 0.38 | 2.89 | 0.05 | 2.91 | 0.10 | 2.89 | 0 |

As demonstrated in previous work,[25] the electronic properties of 1T MXene-$Mo_2C$ are strongly correlated with surface terminations. For example, bare $Mo_2C$ and $Mo_2CO_2$ are metals, while $Mo_2C(OH)_2$ are semiconductors with band gap of 0.1 eV.[25] Unlike the 1T structures, 2H MXene-$Mo_2C$ monolayers are quite different. Fig. 2 shows the band structures of the most stable 2H configurations. Unlike the 1T structures, all of them are metallic, which happens to meet the prerequisite for superconductivity. Note that the shown dependence of electronic properties on different phases could potentially be used to modulate MXene materials, which seems worth exploring further. In general, there are several bands crossing the Fermi level, indicating substantial density of states (DOS) beneficial for higher $T_c$. By further analyzing the projected electronic density of states (PDOS), we noted that DOS near the Fermi level is mainly contributed by the 4d orbitals of the molybdenum atoms.

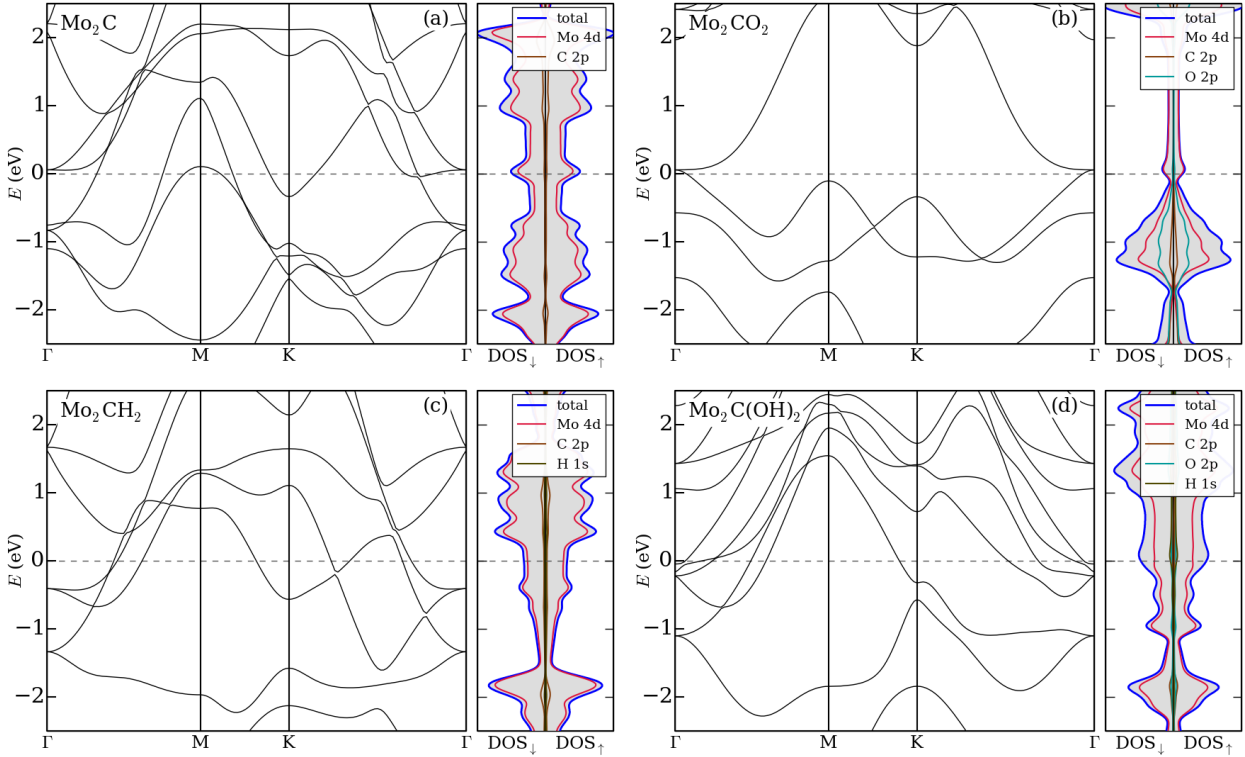

**Fig. 2** Band structures and electronic density of states of (a) $Mo_2C$, (b) $Mo_2CO_2$, (c) $Mo_2CH_2$, and (d) $Mo_2C(OH)_2$ MXenes.

Previously, a theoretical study by Zha *et al.*[32] reported no magnetization in 1T phase MXene-$Mo_2C$. In order to probe the possible magnetism in the 2H phase MXene-$Mo_2C$ monolayers, we performed spin-polarized calculations in bare and functionalized structures. As can be seen from the electronic density of states (DOS) in Fig. 2, the majority and minority spin projections are symmetric, clearly indicating no magnetization. Thus, the nonmagnetic nature of $Mo_2C$ material is robust and cannot be altered readily by phase transformation or surface functionalization.

After fully exploring the electronic properties of MXene-$Mo_2C$ and its functionalized derivatives, we now examine the possibility of conventional superconductivity. Due to large computational expense, we only studied the energetically more favorable 2H MXene-$Mo_2C$ and chose $Mo_2CO_2$, $Mo_2CH_2$ and $Mo_2C(OH)_2$ as the representatives for the functionalized derivatives, and only considered the most stable conformers. All calculations were performed with spin-unpolarized systems.

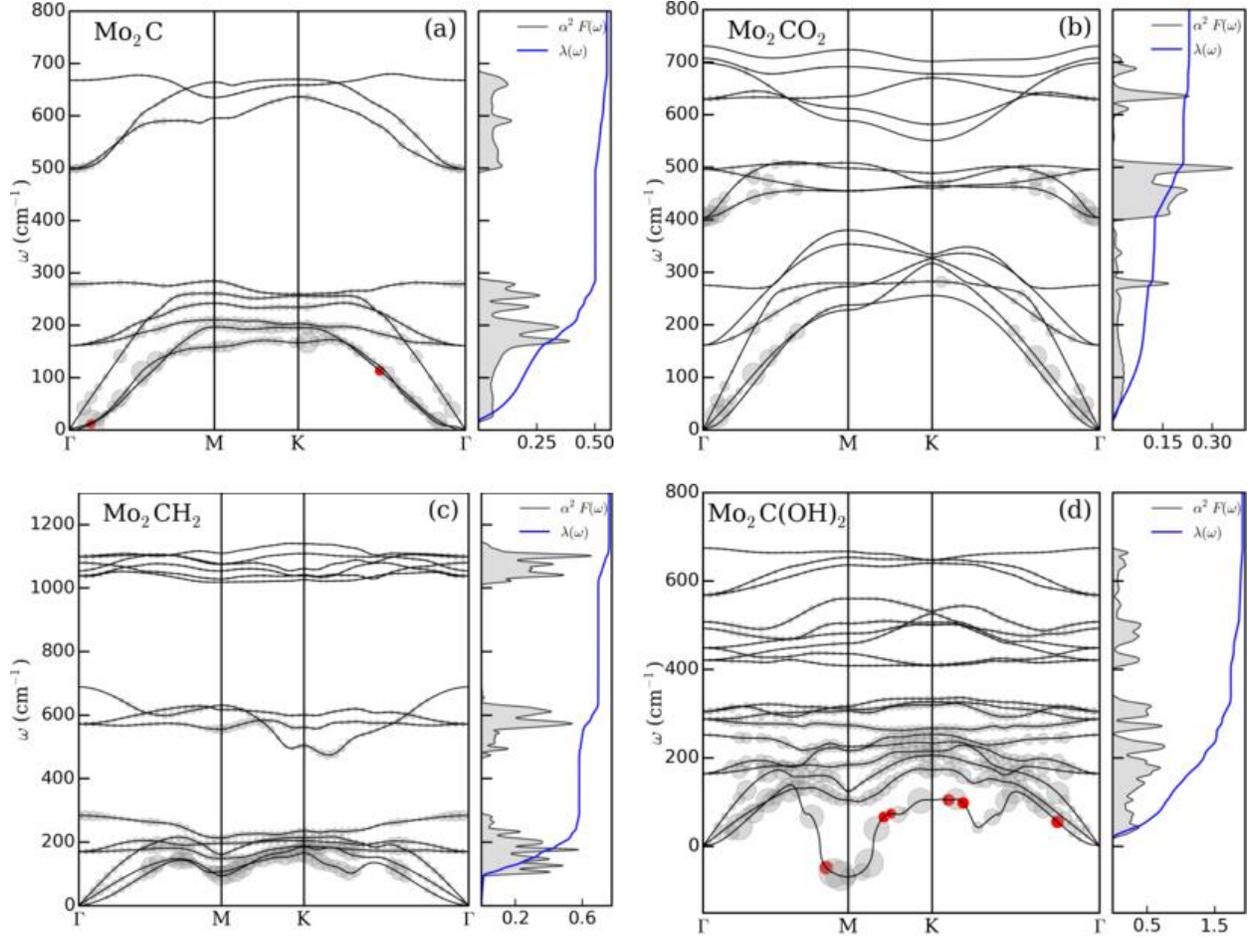

**Fig. 3** Phonon dispersions, electron-phonon interactions, Eliashberg spectral function, and the frequency-dependent electron-phonon couplings of (a) $Mo_2C$, (b) $Mo_2CO_2$, (c) $Mo_2CH_2$, and (d) $Mo_2C(OH)_2$. A small strain of 1% has been applied to the $Mo_2CH_2$ structure. The areas of the grey and red circles are proportional to the magnitude of the electron-phonon interaction. The areas of the red circles have been reduced 10-fold.

Fig. 3 shows the phonon spectra, electron-phonon interactions, Eliashberg spectral function, and the frequency-dependent electron-phonon coupling parameter $\lambda(\omega)$ of bare and functionalized 2H MXene-$Mo_2C$. The phonon spectra are characterized by distinct sets of well-separated frequency bands. The first band lies below 300 cm$^{-1}$, and consists of the three acoustic modes and Mo optical modes. The acoustic modes near $\Gamma$ point display two linear in-plane and one parabolic out-of-plane dispersion curves, characteristic of a 2D system. Between the two in-plane linear acoustic modes, the lower-energy one is longitudinal, and the higher-energy one is transverse. Accordingly, along $\Gamma\rightarrow M$, the lower-energy in-plane acoustic mode is polarized in the armchair direction, whereas the higher energy mode is polarized in the zigzag direction. The first two optical modes are doubly degenerate at the $\Gamma$ point, and represent the in-plane twisting motion of the Mo atoms. The third optical mode is the transverse stretching vibration of the $Mo_2$ dimer. Remarkably, these low-frequency modes are present and have similar energies in all of the structures. The second band lies between 300 cm$^{-1}$ and 800 cm$^{-1}$ and mainly corresponds to the in-plane (lower branches) and out-of-plane (higher branches) optical vibrations of C and O atoms. For hydrogen-containing structures, there is also a band above 800 cm$^{-1}$ (see Fig. S2 for Mo$_2$C(OH)$_2$) due to the vibrations of H atoms and/or O-H bond.

The contributions of different phonon bands can be seen from the electron-phonon coupling parameter $\lambda(\omega)$. The first phonon band generally gives the greatest contribution to the total electron-phonon coupling. As can be seen from Fig. 3, the contribution of the band to the total coupling $\lambda$ varies significantly among different monolayer structures: it is reduced from 0.50 in Mo$_2$C to 0.12 in Mo$_2$CO$_2$, whereas it increases to 0.56 in Mo$_2$CH$_2$, and to 1.53 in Mo$_2$C(OH)$_2$. This indicates that functionalization can be used to effectively tune the coupling. The contribution of the second band increases from 0.05 in Mo$_2$C to 0.11 in Mo$_2$CO$_2$ and Mo$_2$CH$_2$, to 0.18 in Mo$_2$C(OH)$_2$, but remains small overall. The third band gives small $\lambda$ contributions of 0.07 and 0.01 in Mo$_2$CH$_2$ and Mo$_2$C(OH)$_2$ respectively.

The flexural mode of the Mo$_2$CH$_2$ structure has very small imaginary frequencies near the Γ point (see Fig. S2). The small imaginary frequency could be an artifact of the simulation, or reflect the actual lattice instability, in this case, the large-wave undulations of 2D materials. This type of instability has been previously seen in 2D boron structures, and can be remedied by applying slight strain or using substrate.[6] Accordingly, we applied a 1% strain to the Mo$_2$CH$_2$ structure and successfully eliminated the small imaginary frequencies. In contrast, the lowest phonon mode of the Mo$_2$C(OH)$_2$ structure has large imaginary frequencies around the M point, as seen in Fig. 3d, indicating structural instability and possible superstructure. Structural reconstruction has been confirmed from a calculation with a 2×2 supercell (See Fig. S3). We also performed phonon calculations in a 2×2 supercell of Mo$_2$C(OH)$_2$. Unfortunately, imaginary phonon frequencies still occur, possibly pointing to a continuous range of wavelengths at which Mo$_2$C(OH)$_2$ is unstable, consistent with a large region of imaginary frequencies seen around the M point in Fig. 3d. We thus regard Mo$_2$C(OH)$_2$ as being structurally unstable.

We have calculated the superconducting critical temperatures $T_c$, as given by the McMillan equation.[33] The obtained critical temperatures are 3.2 K for Mo$_2$C, ~0 K for Mo$_2$CO$_2$, and 12.6 K for Mo$_2$CH$_2$. Although unstable, we still formally evaluated the $T_c$ for the Mo$_2$C(OH)$_2$ structure. The obtained $T_c$=25.5 K for Mo$_2$C(OH)$_2$ is surprisingly high, although it is unclear whether the lattice instability drives the $T_c$ up in this structure. The increase in $T_c$ in hydrogen-containing structures as compared to bare Mo$_2$C is mostly due to the increased electron-phonon coupling, whereas the vanishing $T_c$ for Mo$_2$CO$_2$ structure is a result of the depletion of electronic density of states near the Fermi level, as can be seen in Fig. 2.

In order to gain some insight into the change of $T_c$ from monolayer to bulk, we have additionally performed a calculation of the $T_c$ for the bulk alpha phase Mo$_2$C, obtaining $T_c = 5$ K. This value agrees favorably with the experimentally measured $T_c$ in this material, for which the reported values are in the range between 4 K and 12 K.[8] The calculated $T_c$ of 5 K in the bulk $Mo_2C$ is slightly higher than the calculated value of 3 K in monolayer $Mo_2C$, in agreement with experimental trends for the dependence of the $T_c$ on thickness.[9] In addition, our preliminary results for $Ti_2C$ and $Ti_2CH_2$ MXenes show much weaker electron-phonon coupling as compared with $Mo_2C$ and $Mo_2CH_2$, yielding much lower $T_c$ (~1.3 K and 0 K for $Ti_2C$ and $Ti_2CH_2$, respectively). Further studies are needed to determine whether $Mo_2C$ with its rather large $T_c$ is an exception in the MXene family, or whether tuning by functionalization could lead to even higher $T_c$ in other MXenes.

## Conclusions

In summary, first-principles calculations of monolayer $Mo_2C$ were performed. Among the three considered structures (monolayer alpha-$Mo_2C$, 1T MXene-$Mo_2C$, and 2H MXene-$Mo_2C$), the 2H MXene-$Mo_2C$ was found to be the most stable. Despite possible passivation, the 2H MXene-$Mo_2C$ preserved its lowest-energy status and the metallic state no matter whether the functional groups exist or not. The possibility of conventional superconductivity and the effect of the functional groups on the critical temperature were subsequently explored. The critical temperatures varied between ~0 and 13 K as the termination functional groups changed. The rather high and well-tunable critical temperatures make this new 2D material an appealing superconductor.

## Computational method

All our results were obtained from first-principles density-functional theory (DFT) calculation within the local-density approximation (LDA)[34] using projector-augmented wave method, as implemented in the Quantum ESPRESSO[35] package. The plane-wave cutoff energy was 60 Ry and converged results were obtained using a Monkhorst-Pack grid of 63×63×1 for both $k$-mesh and $q$-mesh. A smearing of 0.01 Ry was used. In order to avoid the interactions generated by the periodic boundary condition, the interlayer distance was set to at least 10 Å. The geometries of all structures were relaxed to the minimum energy configurations by following the forces on atoms and the stress tensor on the unit cell.

The evaluation of the critical temperatures for the superconducting transition was based on the microscopic theory of Bardeen, Cooper, and Schrieffer (BCS),[21] with the rigorous treatment of electron-phonon interactions introduced by Migdal[36] and Eliashberg.[37] Phonon frequencies and electron-phonon coupling coefficients were calculated using density-functional perturbation theory.[38] The actual $T_c$ values reported here were obtained from the analytical approximation given by the McMillan equation,[33] further modified by Allen and Dynes:[39]

$$k_B T_c = \frac{\hbar \omega_{\ln}}{1.20} \exp\left[\frac{-1.04(1+\lambda)}{\lambda - \mu^* - 0.62\lambda\mu^*}\right]$$

The prefactor $\omega_{\ln}$ was the logarithmically averaged phonon frequency and the effective electron-electron repulsion $\mu^*$ was treated as an empirical parameter with the value $\mu^* = 0.1$.[40,41]

# Supplementary information

Electronic supplementary information (ESI) available: Geometries of monolayer $Mo_2C$ obtained by truncating bulk alpha-$Mo_2C$, additional details of phonon dispersions, electron-phonon interactions, Eliashberg spectral function, and the frequency-dependent electron-phonon couplings of $Mo_2CH_2$ and $Mo_2C(OH)_2$, and optimized geometry of the 2×2 $Mo_2C(OH)_2$ superstructure, convergence tests.

# Supplementary Information for: Predicting Stable Phase Monolayer Mo₂C (MXene), a Superconductor with Chemically-Tunable Critical Temperature

Jincheng Lei, Alex Kutana, and Boris I. Yakobson*

*Department of Materials Science and NanoEngineering, Rice University, Houston, Texas 77005*

E-mail: biy@rice.edu

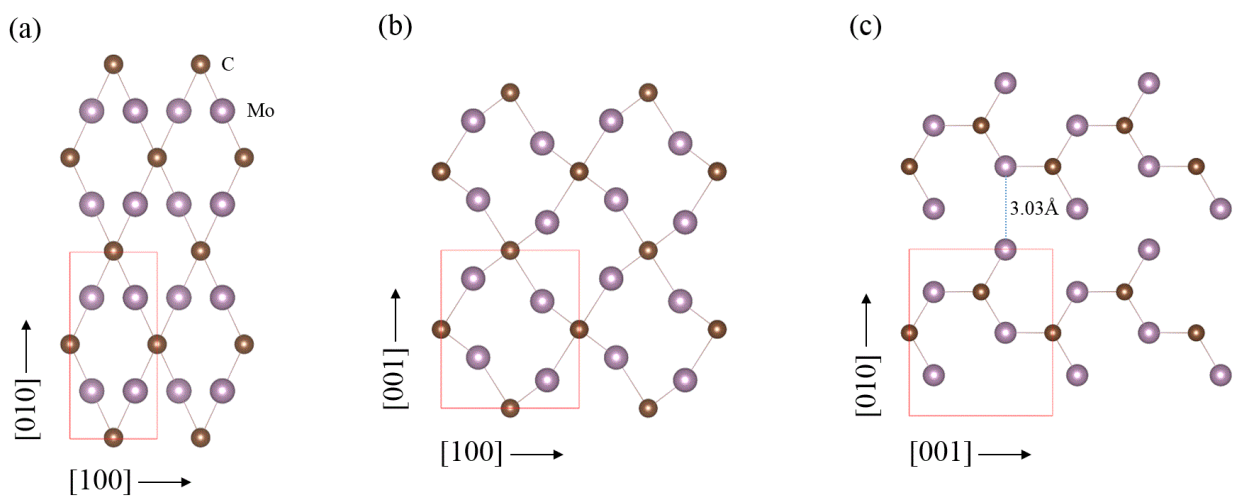

**Fig. S1** Geometries of monolayer Mo$_2$C obtained by truncating bulk alpha-Mo$_2$C: (a) normal to [001] direction, (b) normal to [010] direction, and (**c**) normal to [100] direction. The red solid lines exhibit the unit cells. Structure (a) is lower in energy by 0.3 eV with respect to (b). Mo atoms are not bonded in (**c**).

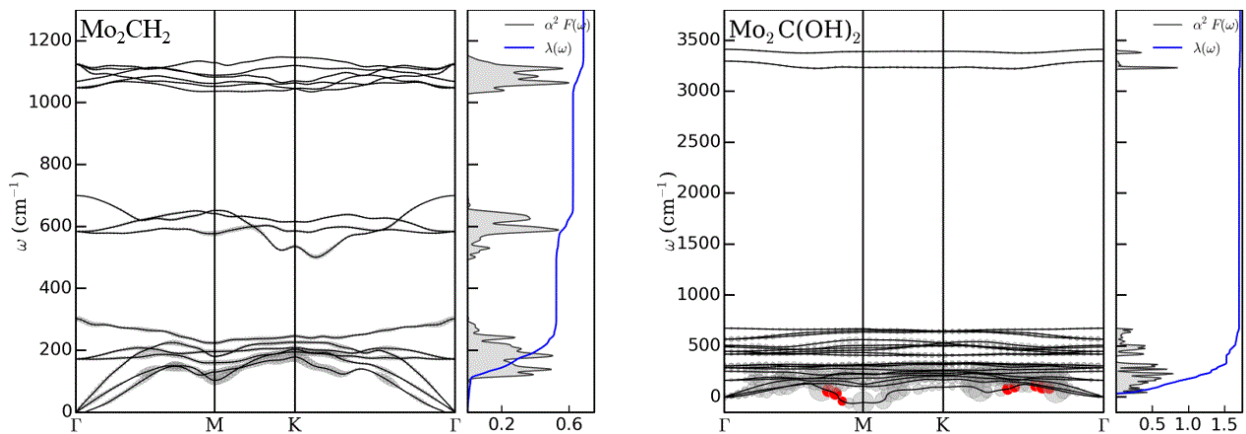

**Fig. S2** Phonon dispersions, electron-phonon interactions, Eliashberg spectral function, and the frequency-dependent electron-phonon couplings of (a) $Mo_2CH_2$, and (b) $Mo_2C(OH)_2$. Here, the q-mesh of 21×21×1 was used for both structures.

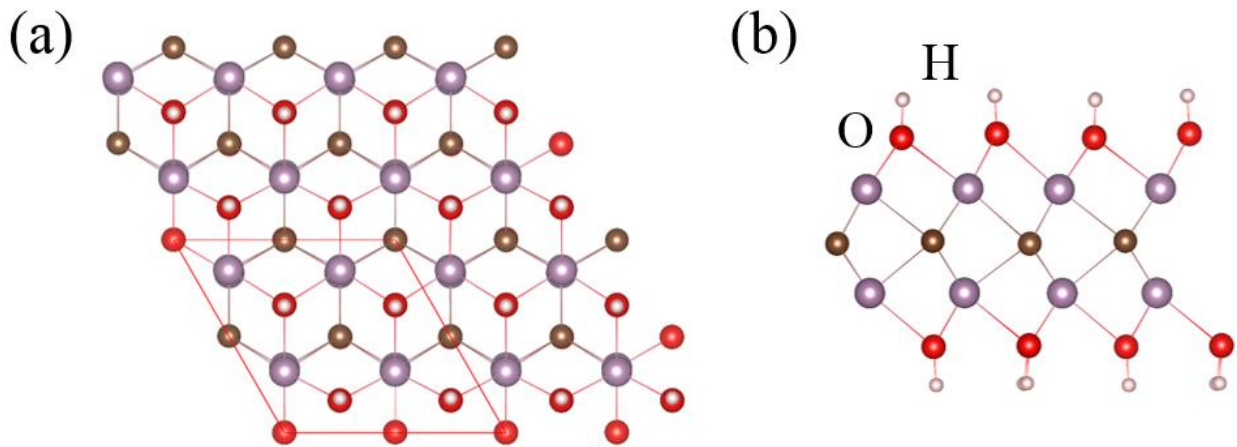

**Fig. S3** Optimized geometry of the 2×2 $Mo_2C(OH)_2$ superstructure: (a) top, and (b) side views. This structure exhibits buckling.

**Table S1.** Convergence test for $Mo_2C$ MXene.

| k-grid | q-grid | $\omega_{\ln}$ [K] | $\lambda$ | $T_c$ [K] |
|---|---|---|---|---|
| 21×21 | 21×21 | 192.151 | 0.53594 | 3.033 |
| 63×63 | 63×63 | 179.701 | 0.55554 | 3.213 |

**Table S2.** Convergence test for $Mo_2CO_2$ MXene.

| k-grid | q-grid | $\omega_{\ln}$ [K] | $\lambda$ | $T_c$ [K] |
|---|---|---|---|---|
| 21×21 | 21×21 | 288.195 | 0.23567 | 0.006 |
| 63×63 | 63×63 | 299.649 | 0.23134 | 0.004 |

**Table S3.** Convergence test for $Mo_2CH_2$ MXene.

| k-grid | q-grid | $\omega_{\ln}$ [K] | $\lambda$ | $T_c$ [K] |
|---|---|---|---|---|
| 21×21 | 21×21 | 329.806 | 0.68663 | 10.936 |
| 63×63 | 63×63 | 326.543 | 0.73205 | 12.626 |

**Table S4.** Convergence test for $Mo_2C(OH)_2$ MXene.

| k-grid | q-grid | $\omega_{\ln}$ [K] | $\lambda$ | $T_c$ [K] |
|---|---|---|---|---|
| 21×21 | 21×21 | 198.947 | 1.71780 | 25.545 |
| 63×63 | 63×63 | 169.780 | 1.92960 | 23.817 |

**Table S5.** Convergence test for $Ti_2C$ MXene.

| k-grid | q-grid | $\omega_{\ln}$ [K] | $\lambda$ | $T_c$ [K] |
|---|---|---|---|---|
| 21×21 | 7×7 | 308.050 | 0.39861 | 1.268 |
| 21×21 | 21×21 | 305.756 | 0.40220 | 1.324 |

**Table S6.** Convergence test for $Ti_2CH_2$ MXene.

| k-grid | q-grid | $\omega_{\ln}$ [K] | $\lambda$ | $T_c$ [K] |
|---|---|---|---|---|
| 21×21 | 7×7 | 540.798 | 0.13062 | 0.000 |
| 21×21 | 21×21 | 530.139 | 0.12895 | 0.000 |